%% file: RAA_4_2.tex
\documentclass{raa}            
\usepackage{graphicx,times}             
\usepackage{subcaption}
\usepackage{booktabs}
\usepackage{graphicx}
\usepackage{natbib}
\usepackage{amssymb,amsmath}
\bibpunct{(}{)}{;}{a}{}{,}

\usepackage[pagebackref=true]{hyperref}

\begin{document}

  \title{Enhancing Galaxy Classification with U-Net Variational Autoencoders. III. Disk-like Galaxy Identification in JWST Samples of up to redshift 4}

   \volnopage{Vol.0 (20xx) No.0, 000-000}      
   \setcounter{page}{1}          

   \author{S. S. Mirzoyan 
      \inst{1,2}
    \and
    A. Avagyan
        \inst{1}
   }

   \institute{Center for Cosmology and Astrophysics, Alikhanyan National Laboratory, 2 Alikhanyan Brothers str., Yerevan 0036, Armenia; {\it mserg@yerphi.am}\\
        \and
             Yerevan State University, Yerevan 0025, Armenia\\
\vs\no
   {\small Received 20xx month day; accepted 20xx month day}}

\abstract{ 
In this third study of the series, we extend our U-Net Variational Autoencoder-based 
galaxy classification framework to a significantly larger JWST sample spanning 
the redshift range $0.5 < z < 4$. Focusing on massive systems with stellar masses 
exceeding $10^{10}\,M_\odot$, we analyze 1{,}380 galaxies that satisfy these criteria 
and apply our previously developed denoising and classification pipeline to identify 
disk-like morphologies across cosmic time. Within this population, our classifier 
detects 382 disk-like galaxies, with a subset showing uncertain features consistent 
with the expected performance limits of current deep-learning models.
This expanded dataset allows us to examine the distribution of disk-like systems in 
a statistically meaningful high-redshift regime, including epochs where well-ordered 
disks are traditionally expected to be rare. The results demonstrate that disk-like 
structures persist across a broad range of redshifts and stellar masses, suggesting 
that massive disks may be more common in the early universe than previously assumed. 
These findings emphasize the value of combining advanced denoising techniques with 
machine-learning-based morphological analysis for characterizing galaxy populations 
in large JWST surveys.
}

   \authorrunning{S. S. Mirzoyan} 
   \titlerunning{JWST High Redshift Galaxy Classification}  

   \maketitle
%
\section{Introduction}           
\label{sec:intro}
Understanding when and how ordered disks assemble and survive in massive galaxies requires a uniform, redshift-spanning morphological census anchored to astrophysically motivated selection criteria. Classical frameworks emphasize the diversity of morphologies within broad Hubble classes \citep{buta2019systematics, mo2010galaxy, sparke2000galaxies, sandage2005classification}, while comprehensive reviews highlight the physical drivers of morphological transformation across cosmic time \citep{bell2026low, conselice2014evolution, naab2017theoretical, genel2018size}. Robust inferences at high redshift are sensitive to surface-brightness dimming, PSF and bandpass effects, and morphology-$k$-corrections \citep{ conselice2024epochs, conselice2023discovery, conselice2020cosmic, lotz2004new, mager2018galaxy, van20143d}. A mass-limited strategy mitigates some of these biases and enables direct comparison across epochs.

\textit{JWST} (James Webb Space Telescope) \citep{gardner2006jwst, rigby2023early} imaging extends quantitative morphology well into the early universe, with large public catalogs facilitating population-level analyses \citep{genin2025dawn, carreira2026jwst, shuntov2025cosmos2025, sun2024structure}. Spectroscopic surveys further contextualize structural evolution during reionization and Cosmic Noon \citep{meyer2025jwst, belli2025blue, rieke2023data, van20143d}. These data require morphological pipelines that can handle heterogeneous signal-to-noise conditions, diverse structural features, and domain-shift effects across filters and redshifts.

Recent \textit{JWST}-based observational studies have provided important context for the present work. In particular, visual classifications from the CEERS (Cosmic Evolution Early Release Science) survey have revealed a surprisingly high fraction of spiral-like galaxies at intermediate redshifts, with up to nearly half of systems at $z \sim 0.5$ exhibiting discernible spiral structure, declining to $\sim 8\%$ by $z \sim 4$ \citep{kuhn2024jwst}. These fractions exceed earlier estimates based on \textit{HST} imaging and suggest that dynamically cold disks and spiral structure may have emerged at much earlier cosmic times than traditionally assumed. Artificial redshifting tests further indicate that intrinsic spiral fractions below $\sim 10$--$20\%$ are statistically disfavored at $z \sim 2$--$3$, motivating mass-limited and resolution-aware morphology analyzes when probing disk formation in the early universe. Independent structural analyses based on \textit{JWST}/NIRCam surface-brightness decompositions have further reported that flattened, disk-like components are already common at high redshift, with only a modest decline in disk-dominated systems toward earlier epochs for massive galaxies \citep{tsukui2026disk}. Broader morphology surveys spanning multiple public \textit{JWST} fields have likewise reported high disk fractions at low and intermediate stellar masses, while finding a pronounced decline in disk dominance at the highest masses, where spheroid-like morphologies become increasingly prevalent \citep{lee2024morphology, carreira2026jwst}. Extending disk studies to even earlier epochs, \citet{mirzoyan2026highz} found that disk-like morphologies constitute a significant fraction of the population at $4 \lesssim z \lesssim 7.7$, emphasizing the importance of noise suppression for robust high-redshift inference.

In parallel, new unsupervised machine-learning approaches have improved the efficiency of large-scale morphology analysis, such as the ConvNeXt+UMAP framework, which combines hierarchical CNN Convolutional Neural Network feature extraction based on the ConvNeXt (Convolutional Neural Network for the 2020s) architecture \citep{liu2022convnext} with nonlinear manifold learning using UMAP (Uniform Manifold Approximation and Projection) \citep{mcinnes2018umap} to classify nearly $10^5$ COSMOS (Cosmic Evolution Survey) galaxies \citep{scoville2007cosmos} into physically meaningful morphology groups using an optimized set of 20 clusters \citep{fang2026updated}. More generally, representation-learning approaches applied to \textit{JWST} data suggest that a fraction of visually identified disks at $z \gtrsim 3$ may occupy a more structurally ambiguous regime when compared against simulated galaxy populations, highlighting potential systematics in visually driven or supervised classifications \citep{vegaferrero2024nature}. Large-scale unsupervised clustering analyses further demonstrate that disk-like galaxies populate multiple distinct regions of morphology space rather than forming a single homogeneous class, underscoring the diversity of disk structures and the sensitivity of disk identification to the adopted representation and clustering strategy \citep{fang2026updated}. Together, these developments highlight both the rapid evolution of galaxy structures across cosmic time and the growing need for automated, noise-robust classification methods capable of operating across heterogeneous imaging datasets.


Machine learning (ML) has become central to modern galaxy morphology studies, lens discovery, and classification pipelines \citep{mirzoyan2025a,  mirzoyan2025b, mirzoyan2019machine, ghaderi2025galaxy, tian2025automatic, atemkeng2025benchmark}. CNN and group-equivariant architectures have shown excellent performance on benchmark galaxy datasets \citep{pandya20232}, though input noise and imaging artifacts continue to limit accuracy—especially for high-redshift sources.

To address this limitation, in Paper~I of this series \citealp{mirzoyan2025a} we introduced a denoising framework based on a U-Net Variational Autoencoder (VAE) \citep{ronneberger2015unet, kingma2013autoencoding}. Trained on realistically contaminated EFIGI (Extraction de Formes Idéalisées de Galaxies en Imagerie) images \citep{baillard2011efigi}, the model effectively suppressed noise while preserving morphology-relevant structures, yielding consistent gains in downstream classification metrics across multiple deep-learning architectures. Paper~I established the methodological foundation that denoising is not merely cosmetic, but functionally beneficial for morphological inference.

In Paper~II \citealp{mirzoyan2025b}, we extended this denoising pipeline to \textit{JWST} Near-Infrared Camera (NIRCam) \citep{rieke2005nircam, rieke2023nircam} observations up to $z \approx 4$. There we demonstrated that VAE-based denoising improves the recoverability of faint structural components and enhances disk/non-disk separability in early-universe galaxies. Analysis of a pilot subsample revealed that disk-like morphologies could be reliably identified even at high redshift, reinforcing the need for denoising-informed classification in low-S/N regimes.

In the present Paper~III, we analyze a mass-selected sample of \textit{JWST} galaxies with $M_\star > 10^{10}\,M_\odot$ over $0.5 < z < 4$. 
Source selection is based on established CANDELS catalogs \citep{CANDELS} to ensure robust stellar-mass and redshift estimates, while all morphological measurements are performed exclusively on \textit{JWST}/NIRCam imaging \citep{CEERS}. 
The construction of the sample and the motivation for this selection are described in detail in Section~\ref{sec:dataset}.

Relative to our earlier papers, the focus in this work is on achieving a more complete sample,  ensuring methodological stability at large scale, and implementing rigorous controls for  systematic effects. Specifically, we (i) reduce domain-shift problems by incorporating  denoising-informed data augmentation and calibrated classification scores and (ii) propagate classification uncertainties by T-scaling.


This paper is structured as follows: Section~\ref{sec:dataset} provides an overview of the dataset employed in our research. In Section ~\ref{sec:method}, we elaborate on the methodologies utilized for data processing and galaxy classification . Section ~\ref{sec:analysis} presents the analysis results and its significance. Lastly, Section ~\ref{sec:conclusion} concludes the paper with a summary of our findings.

\section{Dataset}
\label{sec:dataset}

For the \textit{JWST} data used in this paper, we adopt a filtering approach informed by recent morphological studies that emphasize the connection between structural features, stellar mass, and redshift. 
The \textit{JWST} imaging analyzed in this work is drawn from the Cosmic Evolution Early Release Science (CEERS) survey, an Early Release Science (ERS) program designed to showcase the capabilities of \textit{JWST} for studying galaxy formation and evolution across cosmic time. CEERS targets the EGS field, covering 100 $\mathrm{arcmin}^2$ during Cycle 1 and provides deep, multi-band NIRCam imaging spanning approximately $0.9$--$5\,\mu$m, enabling rest-frame optical morphological studies for galaxies out to $z \sim 4$. 
We utilize the publicly available CEERS NIRCam imaging from the ERS data release (Cycle 1). The images have a native pixel scale of $\sim 0.031^{\prime\prime}$ and achieve a point-spread-function full width at half maximum (FWHM) of approximately $0.06^{\prime\prime}$--$0.09^{\prime\prime}$ across the filters relevant to this study, corresponding to sub-kiloparsec physical resolution over most of the redshift range probed.

To construct the parent galaxy sample, we use the CANDELS photometric catalogs \citep{CANDELS}, which provide homogeneous stellar-mass estimates, photometric redshifts, and extensive ancillary photometry across the EGS field. The CANDELS catalogs are well suited for defining mass-limited samples due to their uniform selection, calibrated stellar population modeling, and established completeness limits. Although \textit{JWST}-based photometric catalogs are now becoming available, we do not use them for parent-sample definition because their mass completeness and cross-field homogeneity are still evolving for large, multi-field samples. In this work, CANDELS is therefore used solely to define the parent sample in stellar mass and redshift, while all morphological measurements are performed exclusively on \textit{JWST}/NIRCam imaging from the CEERS survey. We restrict the parent sample to galaxies with stellar masses $M_\star > 10^{10}\,M_\odot$ and redshifts $0.5 < z < 4$, following the approach of \citet{kuhn2024jwst}. This selection ensures robust stellar-mass completeness and facilitates direct comparison with pre-\textit{JWST} studies.

Because the CEERS \textit{JWST} survey covers only a subset of the full EGS field spanned by the CANDELS catalogs, the CANDELS parent sample is spatially matched to the CEERS footprint prior to retrieving \textit{JWST}/NIRCam imaging. For each galaxy in the selected CANDELS parent sample, we retrieve all available \textit{JWST}/NIRCam observations from the CEERS survey across seven filters: F115W, F150W, F200W, F277W, F356W, F410M, and F444W. This cross-matching procedure yielded a total of 12{,}390 cutout images, corresponding to 1{,}380 unique galaxies. The remaining sources from the initial CANDELS list were not located within the available \textit{JWST} fields. These bands span rest-frame optical to near-infrared wavelengths over $0.5 < z < 4$, allowing different structural components to be enhanced or suppressed depending on wavelength coverage and signal-to-noise conditions.

All \textit{JWST}/NIRCam images are extracted as fixed-size, square cutouts centered on the CANDELS catalog coordinates of each galaxy. Each image stamp has a size of 100$\times$100 pixels, corresponding to an angular extent of approximately $3^{\prime\prime} \times 3^{\prime\prime}$. This field of view is sufficient to capture the full extent of the target galaxies and their immediate surroundings while minimizing background contamination and ensuring consistent input dimensions for the denoising and classification pipeline.

To ensure the integrity of the morphological analysis, all retrieved image cutouts are subjected to an automated quality-control procedure prior to denoising and classification. Images exhibiting extremely low signal levels, abnormally high background noise, incomplete coverage, or near-uniform pixel distributions are identified using statistical diagnostics based on dynamic range, spatial variance, and central-to-background contrast, and are excluded from further analysis. The specific quality-assessment algorithm and rejection criteria are described in detail in Section~\ref{sec:method}.

\begin{figure}[htbp]
    \centering
    \includegraphics[width=0.8\textwidth]{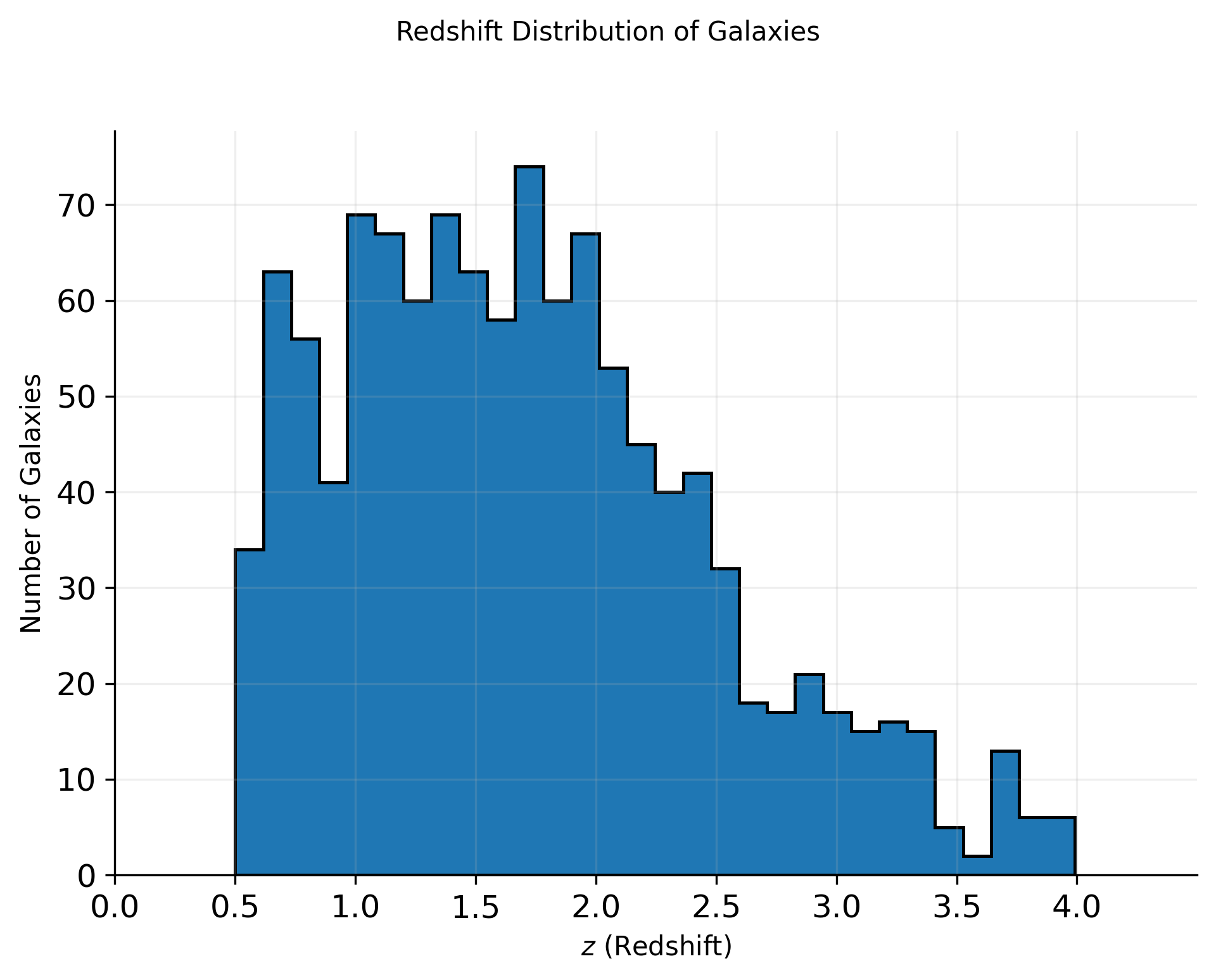}
    \caption{Redshift distribution of the 1{,}380 galaxy sample shown as a histogram, with redshift values plotted along the x-axis and galaxy counts per bin along the y-axis.}
    \label{fig:redshift_distribution}
\end{figure}

After applying the stellar-mass and redshift selection based on the CANDELS catalogs, cross-matching with available CEERS \textit{JWST}/NIRCam imaging, and removing low-quality or unusable frames through automated quality control, our final analysis sample consists of 1{,}380 galaxies with reliable multi-band \textit{JWST} coverage, corresponding to a total of 6{,}730 \textit{JWST}/NIRCam images. These galaxies span the redshift range $0.5 < z < 4$ and have stellar masses $M_\star > 10^{10}\,M_\odot$, forming a mass-complete sample suitable for statistical morphology studies. Application of the denoising and classification pipeline described in Section~\ref{sec:method} identifies 382 galaxies that are flagged as disk-like, with a small additional subset classified as ambiguous. The redshift distribution of the final sample is shown in Figure~\ref{fig:redshift_distribution}.

In addition to the \textit{JWST}/NIRCam imaging from the CEERS survey described above, we employ two auxiliary datasets to train and validate the components of our denoising and morphology–classification pipeline. As part of the methodology, we use the EFIGI dataset exclusively for training and validating the denoising module prior to applying the denoising pipeline to the \textit{JWST} data.

The EFIGI catalog \citep{baillard2011efigi} contains high-quality, visually inspected galaxy images across a wide range of Hubble types. The images are primarily based on optical imaging from surveys such as SDSS, with a typical pixel scale of $\sim 0.4^{\prime\prime}$ per pixel and an effective seeing-limited spatial resolution of approximately $1$--$1.5^{\prime\prime}$. As in Paper~I \citep{mirzoyan2025a}, we use EFIGI to simulate realistic imaging contaminants and validate the performance of the denoising model.

The second dataset used for supervised morphological modeling is the Galaxy10 DECaLS dataset \citep{willett2013galaxy, lintott2011galaxy, walmsley2022galaxy}, which provides curated ten-class morphological labels derived from the Galaxy Zoo project and DECaLS imaging. The DECaLS imaging underlying Galaxy10 is based on ground-based optical observations with a native pixel scale of $0.262^{\prime\prime}$ and a typical point-spread-function FWHM of $\sim 1.0^{\prime\prime}$--$1.3^{\prime\prime}$. As demonstrated in Paper~II, this dataset is well suited for constructing a robust binary disk/non-disk classifier when denoised inputs are supplied to modern deep-learning architectures.

\section{Method}
\label{sec:method}

The methodological framework employed in this work builds upon the general strategy 
introduced in Paper~I of this series, but it is presented here in a fully self-contained form. 
Our objective is to obtain morphology-preserving denoised galaxy images before performing 
a supervised disk/non-disk classification. The overall approach consists of two major 
components: (i) training a U-Net Variational Autoencoder (VAE) to remove realistic 
astrophysical and instrumental contaminants, and (ii) applying a rotation- and 
reflection-equivariant classifier to distinguish disk-like systems from non-disks.

\subsection{Image Quality Assessment and Filtering}

Our objective is to obtain morphology‑preserving denoised galaxy images, following automated image‑level quality assessment, before performing a supervised disk/non‑disk classification.

\begin{figure}[h!]
    \centering

    \begin{minipage}{0.19\textwidth}
        \centering
        \includegraphics[width=\textwidth]{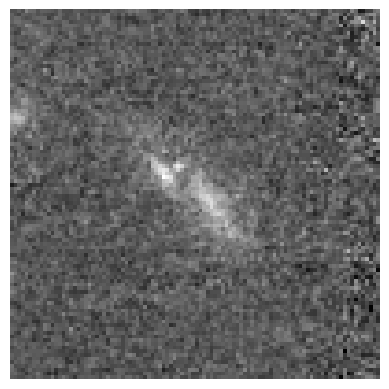}
    \end{minipage}
    \hfill
    \begin{minipage}{0.19\textwidth}
        \centering
        \includegraphics[width=\textwidth]{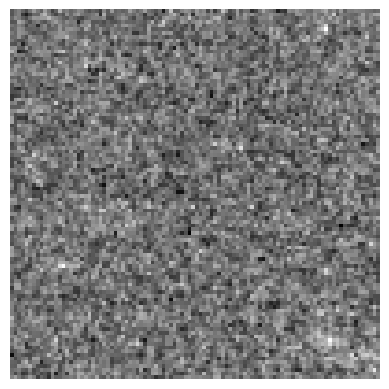}
    \end{minipage}
    \hfill
    \begin{minipage}{0.19\textwidth}
        \centering
        \includegraphics[width=\textwidth]{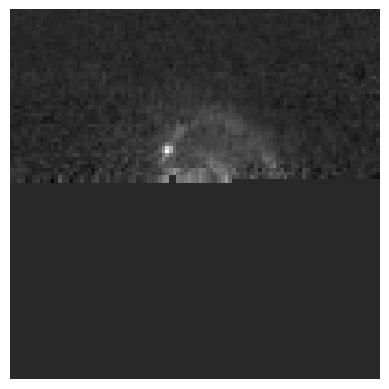}
    \end{minipage}
    \hfill
    \begin{minipage}{0.19\textwidth}
        \centering
        \includegraphics[width=\textwidth]{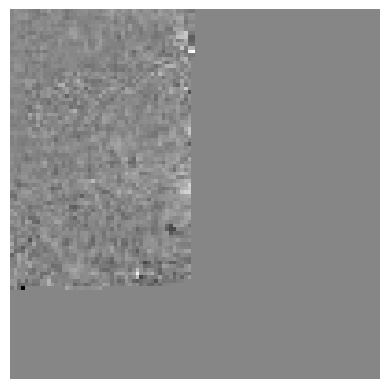}
    \end{minipage}
    \hfill
    \begin{minipage}{0.19\textwidth}
        \centering
        \includegraphics[width=\textwidth]{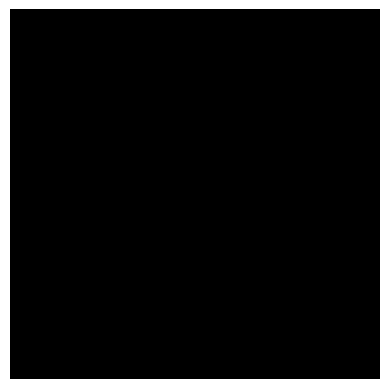}

    \end{minipage}

    \caption{Representative examples of JWST cutout images rejected by the automated quality‑assessment pipeline, including noise‑dominated frames, partially corrupted images, incomplete coverage artifacts, and blank fields lacking reliable morphological information.}
    \label{fig:rejected_frames}
\end{figure}

A non-negligible fraction of the retrieved images exhibit either extremely low signal levels or exceptionally high background noise. In such cases, the classifier’s predictions become unreliable, and these frames must be excluded from the final sample.
To accomplish this, we employ a self-developed, automated quality-assessment algorithm that applies a series of statistical and structural checks to identify and remove low-quality images prior to inference. The quality-assessment algorithm is tailored specifically for automated pre-selection of JWST cutouts before morphological analysis. 
Representative examples of the rejected frames are shown in Figure~\ref{fig:rejected_frames}.

For images passing these global checks, the algorithm compares the statistical properties of a central region to those of an outer annulus. Images are rejected if the central region fails to exhibit sufficient contrast or signal-to-noise relative to the background, or if gradient-based structure in the central region is not enhanced compared to the outskirts. These criteria remove frames with detectable flux but lacking the centralized signal or morphological structure required for reliable GCNN-based classification, ensuring that downstream inference operates only on images containing meaningful and interpretable galaxy information.

The filtering procedure operates on the grayscale representation of each image and evaluates several statistical diagnostics. First, it rejects globally uniform or near-uniform frames by examining their overall standard deviation and dynamic range. It then inspects large spatial regions—halves and quarters of the image—to detect uniform blocks indicative of empty, corrupted, or partially blank exposures. To ensure that the galaxy is not overwhelmed by noise, the algorithm compares the central region with an outer annulus, computing relative contrast, local signal-to-noise ratio, and gradient-based structural differences. Images lacking sufficient central signal or morphological structure—conditions necessary for a reliable GCNN prediction—are discarded.
These statistical checks collectively provide a robust, automated mechanism for separating unusable frames from high-quality galaxy images, ensuring that the subsequent GCNN classifier operates on data where meaningful morphological information is present.

After applying this filtering step, we retain 6{,}730 images corresponding to 1{,}144 galaxies, for which the GCNN model is used to perform the final inference.

\begin{figure}[t]
\centering
\begin{minipage}{0.48\textwidth}
    \centering
    \includegraphics[width=\linewidth]{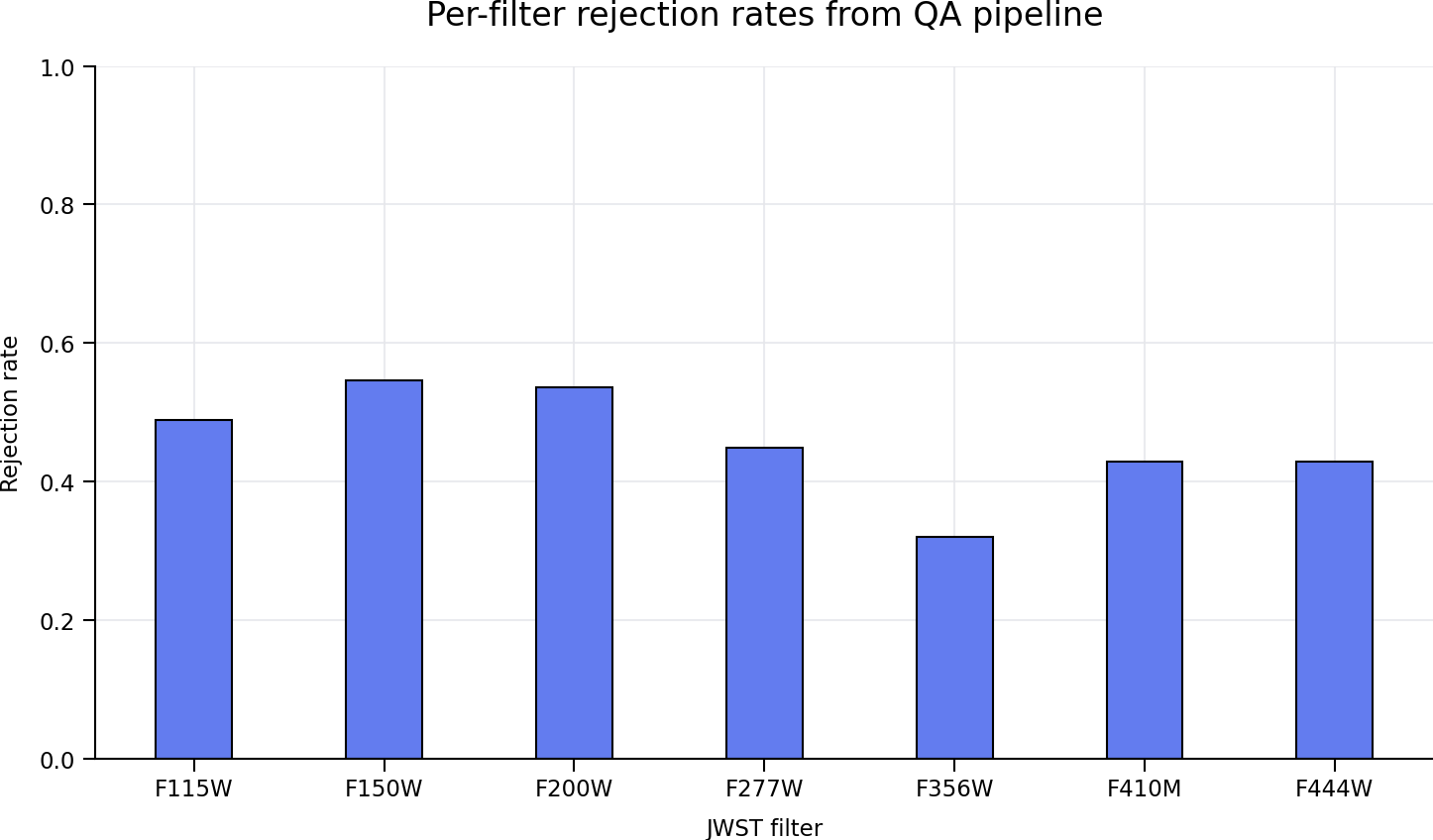}
\end{minipage}\hfill
\begin{minipage}{0.48\textwidth}
    \centering
    \includegraphics[width=\linewidth]{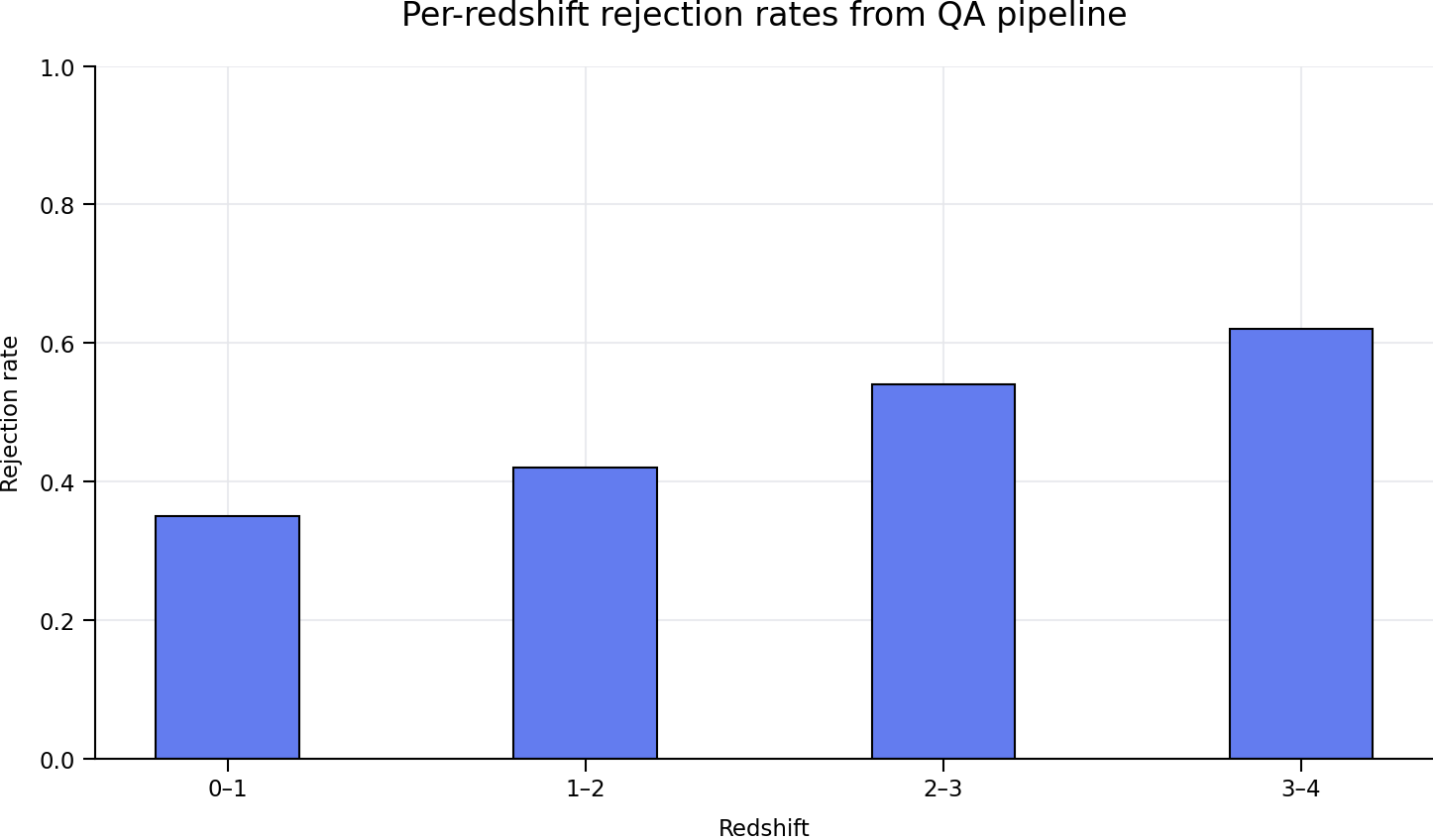}
\end{minipage}

\caption{Image rejection rates from the quality-assessment pipeline.
\textit{Left:} Per-filter rejection rates showing higher
rejection in bluer bands. \textit{Right:} Per-redshift rejection rates shown for diagnostic purposes, illustrating the expected increase toward higher redshift due to
surface-brightness dimming and band-dependent sensitivity.}
\label{fig:qa_rejection_summary}
\end{figure}

Figure~\ref{fig:qa_rejection_summary} summarizes the behavior of the automated quality‑assessment pipeline across observational conditions. The left panel shows the fraction of rejected images as a function of JWST filter, highlighting increased rejection in bluer bands, while the right panel illustrates the redshift dependence of image rejection, demonstrating a systematic increase toward higher redshift consistent with surface‑brightness dimming and band‑dependent sensitivity effects.

\subsection{Training Data and Contamination Modeling}

To construct a suitable training set for the denoising stage, we select a subset of 1,400 visually clean galaxies from the EFIGI survey \citep{baillard2011efigi}.  
These images serve as baseline inputs from which we generate controlled contamination scenarios. Using the \texttt{PyAutoLens} framework \citep{nightingale2023autolens}, each image is augmented by inserting between one and three foreground or background sources—either compact stars or extended galaxies—with randomized structural parameters, light profiles, ellipticities, orientations, and intensities. This process mimics the wide variety of blends, overlaps, and artifacts present in real extragalactic imaging campaigns. Similar contamination simulations have been shown to increase robustness of the model in related studies \citep{yao2024galaxy}.  

From the contaminated dataset, 1000 images are assigned to the training and validation subset, while the remaining 400 are held out for testing. Each clean image undergoes ten independent contamination realizations, yielding a final corpus of 10,000 denoising examples.  
All images are normalized to the $[0,1]$ interval and padded to 
$256 \times 256 \times 3$ pixels so that the output format matches the Galaxy10~DECaLS dataset used later for classification.

\subsection{U-Net VAE Architecture}

The denoising model used in this work is a U-Net–based variational autoencoder (VAE) designed to remove noise while preserving faint morphological structure. The architecture follows a standard encoder–decoder layout: the encoder consists of four convolutional blocks with ReLU activations and max‑pooling operations, which progressively downsample the input image and increase the representation depth. The decoder mirrors this structure using transposed convolutions to restore the original spatial resolution.
Unlike a purely deterministic U‑Net, the model includes a variational latent space from which reconstructions are sampled. During training the VAE is optimized with a composite loss function that balances two objectives: (i) accurate pixel‑level reconstruction of the input image and (ii) regularization of the latent representation so that it remains smooth and well‑behaved. The reconstruction component encourages the network to reproduce the clean underlying structure of the galaxy, while the latent regularization prevents overfitting to noise and ensures that the learned latent space captures statistically meaningful variations in morphology.
A full methodological description, including  used composite loss function  is provided in Paper~II.

\subsection{Binary Disk Identification via Group CNNs}

In Paper~I, we utilized a group-equivariant convolutional neural network (GCNN) model \citep{cohen2016group} to tackle a ten‑class morphological classification problem. In contrast, the current work applies a GCNN architecture to a binary classification task, distinguishing disk-like from non-disk galaxies, following the methodology presented in Paper~II.
As demonstrated in Paper II, this GCNN‑based binary framework performs exceptionally well: trained on the Galaxy10 DECaLS dataset, the model achieved 98.9\% training accuracy, 98.6\% test accuracy, and an F1‑score of 96.5\%, confirming its strong generalization capability and its effectiveness in identifying disk‑like structures even in noisy, high‑redshift images. This reduction to a binary task reflects the scientific goal of identifying rotationally supported structures over a wide redshift range.

For this purpose, we adopt Group Convolutional Neural Networks (GCNNs), which naturally incorporate discrete rotational and reflection symmetries into the network architecture. GCNNs are particularly suited to astronomical imaging, where galaxy orientation on the sky is arbitrary. Following the methodology of \cite{pandya20232}, the classifier is implemented with the \texttt{escnn} library \citep{weiler2021escnn} using D16 symmetry-a group including 8 rotational and 8 reflection operators. This choice ensures equivariance to common geometric transformations without requiring the network to learn them from data, leading to improved performance on both low- and high-redshift samples.

The \textit{JWST} imaging used in this paper extends to redshifts $z \approx 4$, offering a rare opportunity to explore the emergence of disk-like structures near cosmic dawn. All galaxy images are first denoised through the VAE and then passed to the GCNN classifier to generate disk-likelihood scores.

\subsection{Qualitative Validation on JWST Data}

\begin{figure*}
\centering
\includegraphics[width=\textwidth]{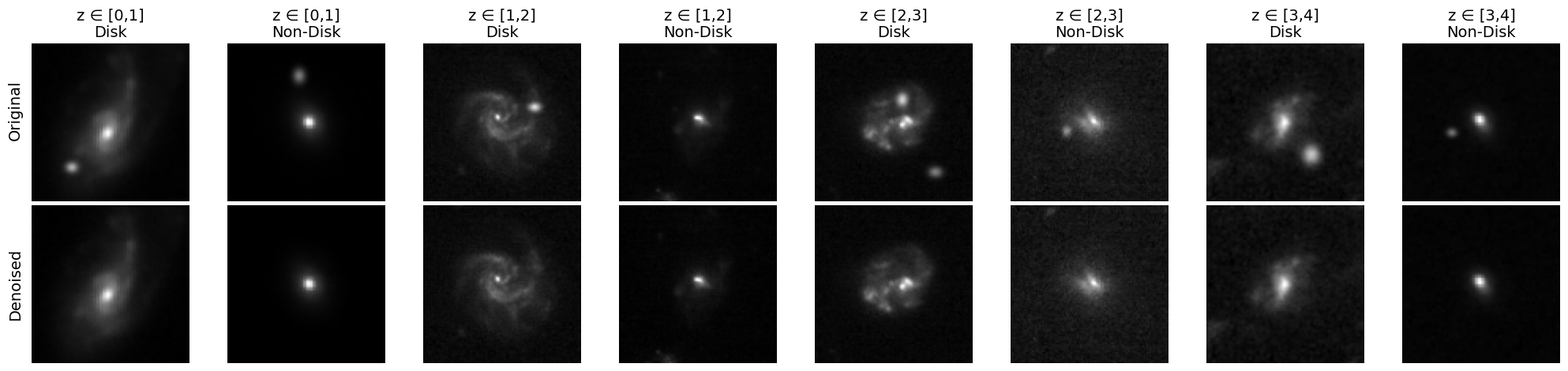}
\caption{Visual validation of the denoising and classification pipeline applied to JWST galaxy images. The top row shows original JWST observations, while the bottom row shows the corresponding denoised
images produced by the U-Net VAE. Each column displays a galaxy classified by the GCNN as either \textit{Disk} or \textit{Non-Disk}, as indicated above the images. Columns 1-2 correspond to galaxies with $z \in [0,1]$, columns 3-4 to $z \in [1,2]$, columns 5-6 to $z \in [2,3]$, and columns 7-8 to $z \in [3,4]$. The denoising process removes contaminating sources while preserving key morphological features, enabling robust disk identification across a wide redshift range despite substantial observational differences between the DECaLS training data and JWST imaging.}
\label{fig:denoise_jwst_examples}
\end{figure*}

A qualitative validation of both the denoising and classification stages is shown in Figure~\ref{fig:denoise_jwst_examples}, using representative JWST galaxy images spanning redshift intervals $z \in [0,1]$, $[1,2]$, $[2,3]$, and $[3,4]$. The top row displays the original JWST observations, while the bottom row shows the corresponding outputs produced 
by the U-Net VAE. In all cases, the denoising model effectively suppresses foreground and background contaminants while preserving the underlying galaxy morphology, including extended low-surface-brightness structure.

Disk and non-disk classifications predicted by the GCNN are indicated above each column. The classifications are consistent across the full redshift range considered. These examples demonstrate that, despite the significant domain shift between ground-based DECaLS imaging used for training and space-based JWST observations, the combined VAE and GCNN framework yields  physically plausible and interpretable disk identification results.

\section{Analysis}
\label{sec:analysis}

The \textit{JWST} imaging used in this study reaches redshifts of $z \approx 4$, providing a rare opportunity to investigate the emergence of disk-like structures near cosmic dawn. All galaxy images are first denoised using the VAE model and subsequently passed through the GCNN classifier to obtain disk-likelihood scores. 


The motivation for collecting multi-filter observations of each galaxy is that different filters can either suppress or enhance key morphological features. Certain wavelengths may obscure disk signatures, while others reveal them more clearly. As a result, our inference procedure is inherently multi-instance: every available image of a galaxy is classified independently. The individual predictions are then aggregated at the galaxy level. For example, if a galaxy has seven available images and the classifier identifies six as non-disk and one as disk, the final classification is assigned as disk. This reflects the physical expectation that genuine disk structures may be detectable only in particular filters, and a single confident disk detection is sufficient to label the system as disk-like.

\subsection{GCNN Inference: Galaxy-level Statistics and High-redshift Disk Fraction}
\label{subsec:keyfindings}
Raw GCNN output probabilities tend to be overconfident for atypical or low-S/N inputs. To correct for this, we adopt a calibration scheme inspired by standard reliability analysis used in modern deep-learning classification tasks. First, we construct a validation subset consisting of denoised Galaxy10 DECaLS images together with their true binary labels. The GCNN is evaluated on this set, and the resulting predicted probabilities are compared to empirical class frequencies in fixed probability bins. We then apply a temperature-scaling transformation, a standard post-hoc calibration technique introduced by \citet{guo2017calibration},

\begin{equation}
\hat{p} = \sigma\left(\frac{z}{T}\right),
\end{equation}

where $z$ is the pre-softmax logit, $\sigma$ is the sigmoid function, and $T$ is a calibration temperature optimized by minimizing the negative log-likelihood on the validation data. This adjustment preserves the model's decision boundaries but rescales probability values to be better aligned with observed classification 
uncertainty. The resulting calibrated scores form the basis for the disk likelihood values used throughout the analysis.

We aggregate image-level predictions across available filters to the galaxy level by grouping sources by identifier and applying an “any-positive” rule: a galaxy is labeled disk-like if at least one of its images is classified as disk. For a continuous confidence metric, we combine per-image probabilities via a noisy-OR aggregator, 
\[
p_{\mathrm{gal}} = 1 - \prod_i (1 - p_i),
\]

which matches the physical expectation that disk signatures may be preferentially revealed in specific bands.

Across the post-filtered sample of \textbf{1,144} galaxies, the overall disk fraction is
\textbf{$f_{\rm disk} = 0.334$} with a Wilson 95\% confidence interval of \textbf{0.307-0.362}.
The distribution of galaxy-level disk probabilities is strongly skewed toward high confidence, with a median
$p_{\rm gal} \approx 1$.

\begin{figure}[t]
    \centering
    \includegraphics[width=0.58\linewidth]{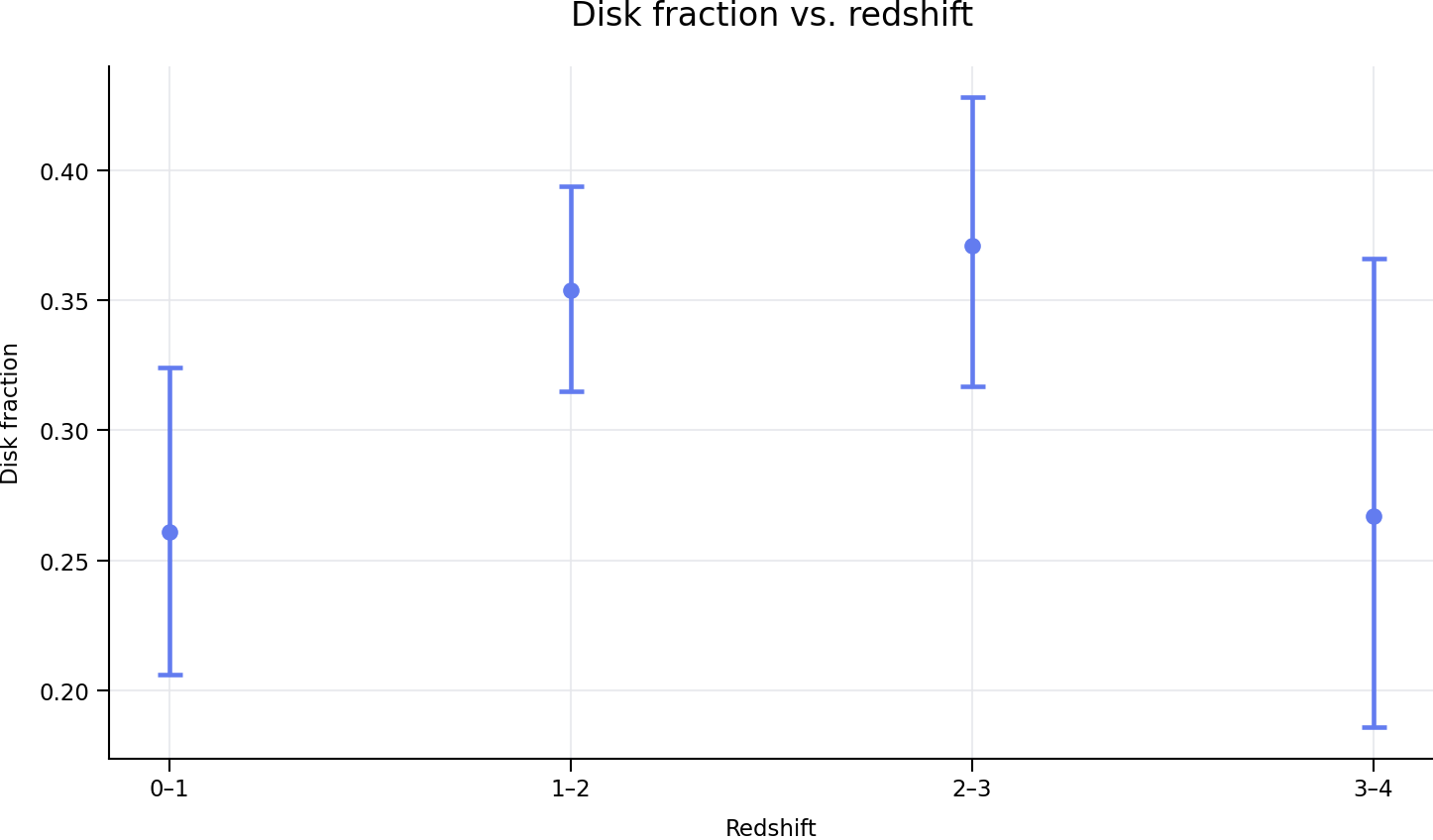}
    \caption{Disk fraction versus redshift bin with Wilson 95\% confidence intervals.}
    \label{fig:diskfrac_z}
\end{figure}

\paragraph{Redshift dependence.}
Partitioning the sample into $z\!\in\![0,1)$, $[1,2)$, $[2,3)$, and $[3,4]$, the disk fractions with Wilson intervals are summarized in Table~\ref{tab:zsummary} and Figure~\ref{fig:diskfrac_z}. 
In the high-redshift regime ($z \ge 3$), we measure a disk fraction of \textbf{0.267} (95\% CI \textbf{0.186-0.366}) based on \textbf{90} galaxies, with a median $p_{\rm gal}$ near unity. These results indicate that rotationally supported systems persist to $z \sim 4$ within our sample.

\begin{table}[t]
    \centering
    \caption{Redshift-binned disk fractions (Wilson 95\% CIs).}
    \label{tab:zsummary}
    \vspace{0.25em}
    \input{z_summary.tex}
\end{table}

\begin{figure}[t]
    \centering
    \includegraphics[width=0.58\linewidth]{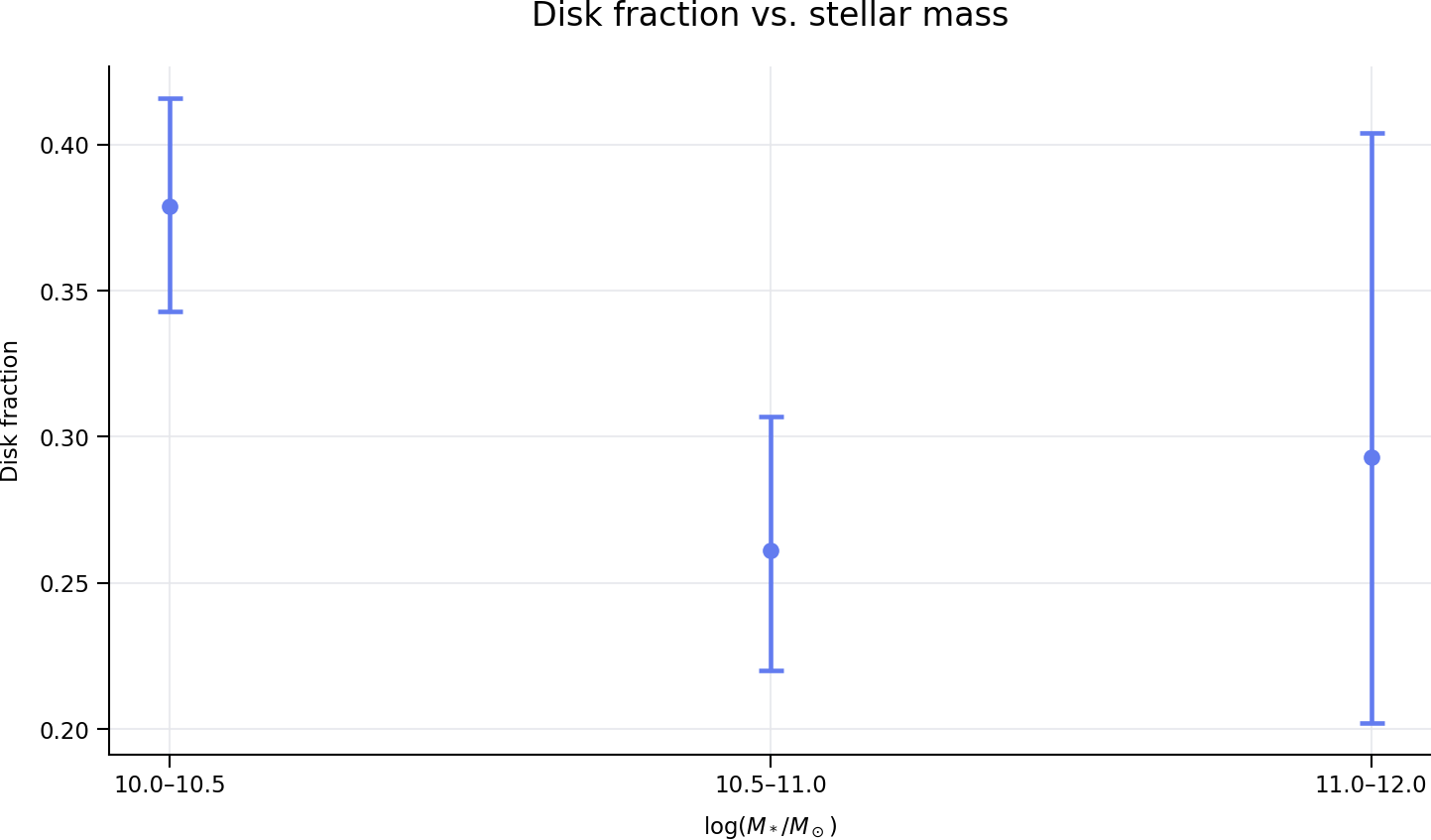}
    \caption{Disk fraction versus stellar mass bin with Wilson 95\% confidence intervals.}
    \label{fig:diskfrac_mass}
\end{figure}

\paragraph{Stellar mass dependence.}
Using stellar-mass bins $\log(M_\star/M_\odot)\in[10.0,10.5)$, $[10.5,11.0)$ and $[11.0,12)$ the corresponding disk fractions are \textbf{0.379} (95\% CI \textbf{0.343-0.416}), \textbf{0.261} (95\% CI \textbf{0.220-0.307}), and \textbf{0.293} (95\% CI \textbf{0.202-0.404}), respectively (Table~\ref{tab:msummary}; Figure~\ref{fig:diskfrac_mass}). Within our sample, the disk fraction peaks around $\log M_\star \sim 10.0$-10.5 and is comparatively lower toward higher masses.

\paragraph{Multi-filter consistency and band effects.}
A total of \textbf{365} galaxies exhibit mixed per-filter labels (both disk and non-disk), underscoring that some bandpasses can suppress disk signatures while others reveal them. At the image level, the mean predicted probability and the fraction of disk classifications vary with filter (Table~\ref{tab:filterdiag}), indicating band-dependent sensitivity that may reflect PSF, S/N, and rest-frame coverage effects.

\paragraph{Comparison with previous studies.}

The disk fractions derived in this work are broadly consistent with recent observational results based on \textit{JWST} imaging, while reflecting differences in sample definition, classification methodology, and selection criteria. Visual classifications from the CEERS survey suggest that the fraction of spiral or disk-like systems declines from $\sim 40$-50\%at $z \sim 0.5$ to $\sim 5$-10\% by $z \sim 4$ \citep{kuhn2024jwst}. Within our mass-limited sample ($M_\star > 10^{10}\,M_\odot$), we measure a disk fraction of $\sim 26$-37\% over $0 < z < 3$, decreasing to $27^{+10}_{-8}\%$ at $z \ge 3$. The comparatively higher fractions
reported here are expected given our focus on massive galaxies, for which disks are both easier to detect and more likely to survive dynamical heating.

Our results are also in close quantitative agreement with recent machine‑learning–based morphology studies at even earlier cosmic times. Using a denoising VAE combined with a rotation‑equivariant GCNN classifier applied to \textit{JWST}/NIRCam imaging at $4 \lesssim z \lesssim 7.7$, \citet{mirzoyan2026highz} reported a disk‑like fraction of $\simeq 0.34$, remarkably similar to the overall disk fraction measured in the present work despite the higher redshift range probed. This consistency suggests that a substantial population of rotationally supported systems is already in place well before $z\sim4$, provided that noise suppression and representation‑aware classification techniques are employed.

Broad \textit{JWST} morphology surveys spanning multiple public fields further support this picture. In particular, \citet{carreira2026jwst} find high disk fractions at low and intermediate stellar masses, with a pronounced decline toward the highest masses where spheroid‑dominated morphologies become increasingly prevalent. This mass dependence is qualitatively consistent with the trends observed in our mass‑limited sample, reinforcing the conclusion that disk survivability decreases at the highest stellar masses, likely due to enhanced merger and compaction activity.

Differences with visual spiral fractions are also attributable to classification definitions: our binary disk/non-disk framework encompasses rotationally supported systems that may lack clearly identifiable spiral arms, particularly at high redshift where surface-brightness dimming, clumpiness, and band-dependent effects obscure detailed arm structure. In this sense, our results should be interpreted as constraints on the prevalence of disk-like stellar mass distributions rather than on visually prominent spiral morphology alone. Similar conclusions have been reached in kinematic and structural studies of massive galaxies at $z \gtrsim 2$, which find evidence for dynamically cold disks even when classical spiral structure is absent \citep{genzel2017early, lelli2023cold}.

\begin{table}[t]
    \centering
    \caption{Stellar-mass-binned disk fractions (Wilson 95\% CIs).}
    \label{tab:msummary}
    \vspace{0.25em}
    \input{m_summary.tex}
\end{table}

\begin{table}[t]
    \centering
    \caption{Per-filter image diagnostics: number of images ($N$), mean predicted probability ($\langle p \rangle$), and fraction classified as disk ($f_{\rm disk}$).}
    \label{tab:filterdiag}
    \vspace{0.25em}
    \input{filter_summary.tex}
\end{table}

\section{Conclusions}
\label{sec:conclusion}
In this work, we have conducted a comprehensive, multi–wavelength morphological analysis of 1,144 galaxies with available \textit{JWST}/NIRCam imaging, leveraging a denoising VAE and a graph-based convolutional neural network to identify disk-like structures deep into the epoch of peak galaxy assembly. Our methodology combines rigorous image-quality filtering, multi-instance classification across seven NIRCam filters, and a conservative galaxy-level aggregation scheme designed to capture genuine disk signatures even when they are present in only a subset of observed bands.
Importantly, within the redshift range $0.5 \le z \le 4.0$, we identify \textbf{382 disk galaxies} out of the \textbf{1,144} systems in our final sample.

After removing low-quality frames, the GCNN classifier achieves robust performance on the remaining 6,730 images. At the galaxy level, we measure an overall disk fraction of $f_{\rm disk} = 0.334^{+0.028}_{-0.027}$, indicating that approximately one-third of galaxies in our sample exhibit disk-like morphology. When examined as a function of redshift, the disk fraction increases from $f_{\rm disk} \sim 0.26$ at $z<1$ to a peak of $\sim0.37$ at $z\sim2$-3, before declining modestly to $0.27^{+0.10}_{-0.08}$ at $z \geq 3$. These results confirm that rotationally supported systems are already well established by $z\sim2$ and remain present—at a non-negligible level—into the early Universe at $z\sim4$, consistent with recent dynamical and photometric evidence for early disk formation.
A similar trend emerges with stellar mass: the disk fraction peaks at $\log(M_\star/M_\odot)\sim10.0$–10.5 and declines toward higher masses. This behavior is broadly consistent with the expectation that intermediate-mass systems are most likely to host dynamically cold disks, whereas massive galaxies undergo more frequent mergers or compaction episodes that disrupt disk structure.
An important outcome of our multi-filter approach is the identification of substantial wavelength-dependent variance in disk detectability. Approximately one-third (365) of the galaxies exhibit mixed per-filter classifications, where at least one filter reveals a disk-like signature while others do not. This highlights the complex interplay between morphology and observed wavelength: shorter-wavelength bands (e.g., F115W, F150W) more frequently show high disk-likelihoods, whereas longer-wavelength filters (e.g., F277W, F356W) often suppress or dilute these features, potentially due to differences in PSF size, surface-brightness sensitivity, and rest-frame wavelength coverage. Our findings therefore underscore the necessity of multi-band classification for robust morphological inference at high redshift.

Finally, our results can be placed in the context of recent \textit{JWST}-based studies that have explored the fraction of disk-like galaxies at high redshift using a variety of complementary approaches. 
Visual classifications from the CEERS survey report high spiral and disk fractions at intermediate redshifts, suggesting that ordered disk structures emerge earlier than inferred from \textit{HST}-based analyses \citep{kuhn2024jwst}.
Machine-learning and representation-based studies applied to the same data have emphasized, however, that a subset of visually identified disks may exhibit structurally ambiguous or dynamically hot configurations, highlighting sensitivity to classification methodology and image quality \citep{vegaferrero2024nature, fang2026updated}.
Structural analyses relying on surface-brightness decompositions likewise find that flattened, disk-like components are common at high redshift, with disk dominance declining primarily at the highest stellar masses \citep{tsukui2026disk, lee2024morphology}.
Extending such analyses to even earlier epochs, denoising-informed classification of \textit{JWST}/NIRCam imaging at $4 \lesssim z \lesssim 7.7$ has demonstrated that disk-like systems constitute a substantial fraction of the population, with inferred disk fractions of order $\sim 0.3$ even within the first billion years of cosmic time \citep{mirzoyan2026highz}.
In this broader context, the disk fraction measured here for massive galaxies over $0.5 < z < 4$ is fully consistent with recent \textit{JWST}-based results, while benefiting from a strictly mass-selected sample, uniform \textit{JWST}/NIRCam imaging, and explicit control of noise-driven systematics through denoising-informed classification. These methodological elements help reconcile differences among existing measurements and reinforce the conclusion that disk-like structures are both common and resilient across a wide range of cosmic epochs.

Taken together, these results provide quantitative evidence that disk-like structures are both common and resilient across cosmic time, even when viewed through the diverse and sometimes challenging observational windows provided by \textit{JWST}. The strong wavelength dependence, high incidence of mixed-filter classifications, and redshift evolution observed here point to a promising future in which machine-learning–assisted, multi-band morphological studies will play a central role in charting the emergence of ordered galactic structure from the early Universe to the present day.

\begin{acknowledgements}
{{{\bf We thank the referee for many useful comments.} We acknowledge the use of the EFIGI, Galaxy10 DECaLS datasets and CEERS database, which enabled the development and validation of this research.}}
\end{acknowledgements}


\label{lastpage}

\end{document}

%% file: z_summary.tex
\begin{tabular}{lrrrrrr}
\toprule
$z_{bin}$ & $N$ & $N_{disk}$ & $f_{disk}$ & low & high & $p_{median}$ \\
\midrule
0–1 & 211 & 55 & 0.261 & 0.206 & 0.324 & 1.000 \\
1–2 & 557 & 197 & 0.354 & 0.315 & 0.394 & 1.000 \\
2–3 & 286 & 106 & 0.371 & 0.317 & 0.428 & 1.000 \\
3–4 & 90 & 24 & 0.267 & 0.186 & 0.366 & 1.000 \\
\bottomrule
\end{tabular}

%% file: m_summary.tex
\begin{tabular}{lrrrrr}
\toprule
M$_{bin}$ & $N$ & $N_{disk}$ & $f_{disk}$ & low & high \\
\midrule
10.0-10.5 & 686 & 260 & 0.379 & 0.343 & 0.416 \\
10.5–11.0 & 383 & 100 & 0.261 & 0.220 & 0.307 \\
11.0–12.0 & 75 & 22 & 0.293 & 0.202 & 0.404 \\
\bottomrule
\end{tabular}

%% file: filter_summary.tex
\begin{tabular}{lrrr}
\toprule
$Filter$ & $N$ & $P_{mean}$ & $f_{disk}$ \\
\midrule
F115W & 905 & 0.907 & 0.305 \\
F150W & 803 & 0.899 & 0.320 \\
F200W & 822 & 0.895 & 0.277 \\
F277W & 977 & 0.937 & 0.068 \\
F356W & 1204 & 0.949 & 0.039 \\
F410M & 1010 & 0.952 & 0.029 \\
F444W & 1009 & 0.954 & 0.025 \\
\bottomrule
\end{tabular}